# A needle-based deep-neural-network camera

RUIPENG GUO,[1] SOREN NELSON,[1] AND RAJESH MENON[1,*]

[1]*Department of Electrical & Computer Engineering, University of Utah, Salt Lake City, UT 84112, USA*
*\*rmenon@eng.utah.edu*

**Abstract:** We experimentally demonstrate a camera whose primary optic is a cannula (diameter=0.22mm and length=12.5mm) that acts a lightpipe transporting light intensity from an object plane (35cm away) to its opposite end. Deep neural networks (DNNs) are used to reconstruct color and grayscale images with field of view of $18^0$ and angular resolution of ~$0.4^0$. When trained on images with depth information, the DNN can create depth maps. Finally, we show DNN-based classification of the EMNIST dataset without and with image reconstructions. The former could be useful for imaging with enhanced privacy.

Photography without lenses has a long and interesting history, starting with the pinhole cameras and more recently using what has been referred to as accidental cameras [1]. Optics-free systems for widefield imaging has also been studied with an emphasis on large field-of-view microscopy [2]. We first acknowledge that incoherent imaging without any optics (*i.e.*, with only free-space propagation) has been studied for a long time, [3] and was recognized as belonging to a class of severely ill-posed problems, uniquely challenging for regularized inversion. Nevertheless, this inverse problem can be solved with regularized matrix inversion in some cases [4,5] as well as with machine-learning techniques [5-7]. Machine-learning has been applied to perform classification tasks on raw [6,7] as well as on anthropocentric images [8,9]. A big advantage of these approaches is that they do not rely on coherent illumination (such as those that exploit speckle) [10,11]. Although utilizing coherence can lead to high spatial resolution, in this work, we explicitly avoid comparing to such approaches, as they are not readily applicable to general-purpose photography.

Imaging is the process of information recovery from the imperfect recorded image. Information from an object is lost along the way to the image sensor, where the image is recorded. This loss of information depends strongly upon the optical system. Conventional lenses have shown their supremacy to minimize such information losses. However, there are many situations, where conventional lenses cannot be used. One example of this is in minimally-invasive deep-brain imaging, where the requirement of minimal brain trauma and low information loss in the imaging channel can be mutually exclusive. Microendoscopes such as a surgical cannula have been used to transport light from inside the brain to the outside sensor, and computational methods can be used to reconstruct the details of the image at the distal end of the cannula, a technique we named computational-cannula microscopy (CCM) [12-14]. CCM can be fast as there is no scanning required. It has already demonstrated in-situ [15] and machine-learning enabled 3D microscopy [16,17]. Here, we extend this approach into full color macro-photography.

The basic principle lies on the premise that as long as an optical system has a linear space-variant transfer function (point-spread function), it may be possible to invert this transfer function to recover the intensity information of the object being imaged [2]. Machine learning has been shown to be an effective approach to solve such inverse problems [18]. By collecting sufficient data, one can train machine-learning algorithms to implicitly extract the space-variant point-spread functions and thereby, recover the object information. By definition, the testing data will limit the nature of objects that could be imaged in such a fashion, which is analogous to restricting the solution space (regularizing) with a-priori information. Here, we demonstrate a camera comprised of a cannula coupled with a deep neural network (DNN) that is trained to recover object information in an anthropocentric format. We further show that it is possible to skip image reconstructions and directly perform classification operations on the raw recorded images, which could enable cameras with enhanced privacy as well as low-power consumption. We experimentally characterized the resolution, field of view, and depth of focus of such a camera with a variety of monochrome and color images.

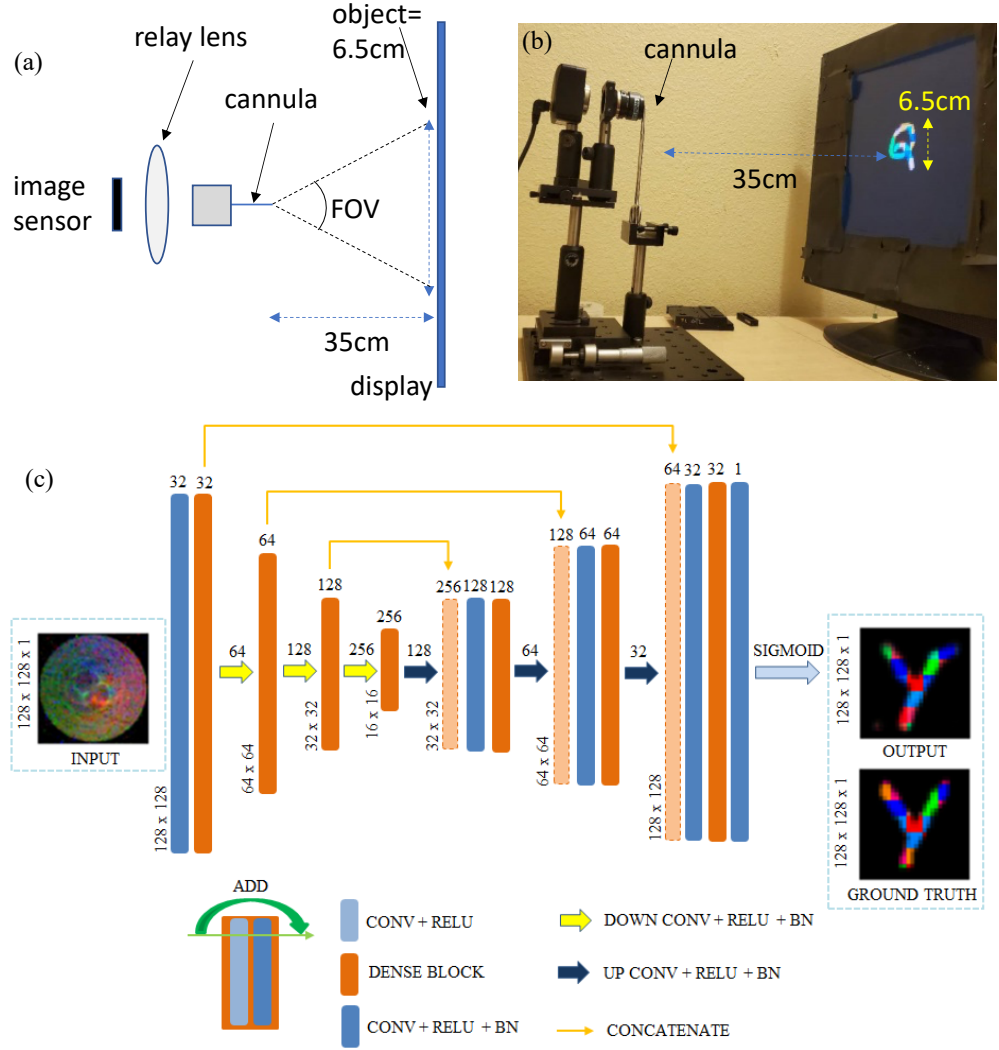

Fig. 1. (a) Schematic and (b) photograph of our experimental setup. (c) Architecture of the modified U-net used for image reconstructions. The operations were performed separately for each color channel. The input is the raw sensor image (example on left) and the output reconstructed and ground truth images (right).

Our experimental setup, described in Fig. 1(a) was comprised of a cannula (diameter=0.22mm, length=12.5mm, Thorlabs CFMC52L02). The distal end of the cannula was facing the object, a liquid-crystal-display (LCD) monitor (ViewSonic VG910b, Size: 19 inch, Resolution: 1280x1024 pixels), placed at a nominal object distance of 35cm from distal end of the cannula. The intensity distribution on the proximal end of the cannula is relayed onto a CMOS image sensor (Amscope MU300, pixel size=3.2μm) via a conventional lens (NAVITAR NMV-12, focal length = 12cm). Different "objects" were displayed on the LCD, and the corresponding images were recorded on the image sensor. We modified the well-known EMNIST dataset [19] by applying random color to the non-zero pixels to create a color EMNIST dataset. 40,000 images from such a modified EMNIST dataset, and 20,890 images from a color emoji dataset [20] were displayed on the LCD and recorded on the sensor (see photograph of system in Fig. 1b). The physical size of each image was 6.5cm square. The size of each recorded image was 160 X 160 pixels (physical size is 0.512mm square) resulting in an effective demagnification of 127x for each side.

Figure 1(c) shows the architecture of our DNN (a modified U-net) that was used for image reconstructions. The raw sensor image (160 X 160 pixels X 3 color channels) was first downsampled to 128 X 128 pixels X 3 color channels as a compromise to keep the DNN small. The U-net is comprised of dense blocks that include two convolutional layers with ReLu activation function and a batch-normalization layer. Pixel-wise cross-entropy was used as the loss function during training. We randomly chose 1000 images from each dataset, and used these exclusively for testing. The structural similarity index (SSIM) and the maximum average error (MAE) were (0.77, 0.04) and (0.67, 0.1) for the color EMNIST and the color emojis, respectively. Exemplary images are shown in Fig. 2, and the EMNIST images are reconstructed with higher fidelity than the emojis. It is noted that this experiment corresponds to a field of view about $2 \times \tan^{-1}(6.5/2/35) = 10.6^0$.

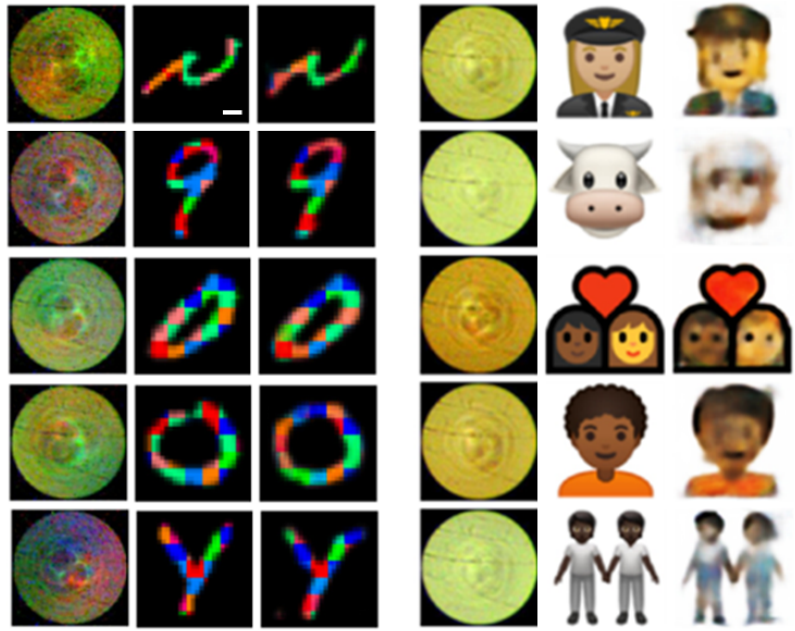

Fig. 2. Color image reconstructions. (a) Color EMNIST images (SSIM=0.77, MAE=0.05) (b) Color emojis (SSIM=0.67, MAE=0.1). Each image is 6.5cm X 6.5cm (white scale bar = 1cm). Left column: raw sensor image. Center column: ground truth. Right column: DNN output. Object distance was 35cm.

In order to estimate the depth of focus (DOF) of our cannula-based camera, we created a new dataset with monochrome and color EMNIST and KANJI [21] characters (40,000 total and 1,000 used exclusively for testing) at 5 distances from 29cm to 41cm. The network was trained for the data at an object distance of 35cm. The averaged SSIM and MAE of the reconstructed images were plotted as function of object distance in Figs. 3a and 3b for grayscale and color images, respectively. This data and the exemplary images shown in Fig. 3c confirm best performance at 35cm.

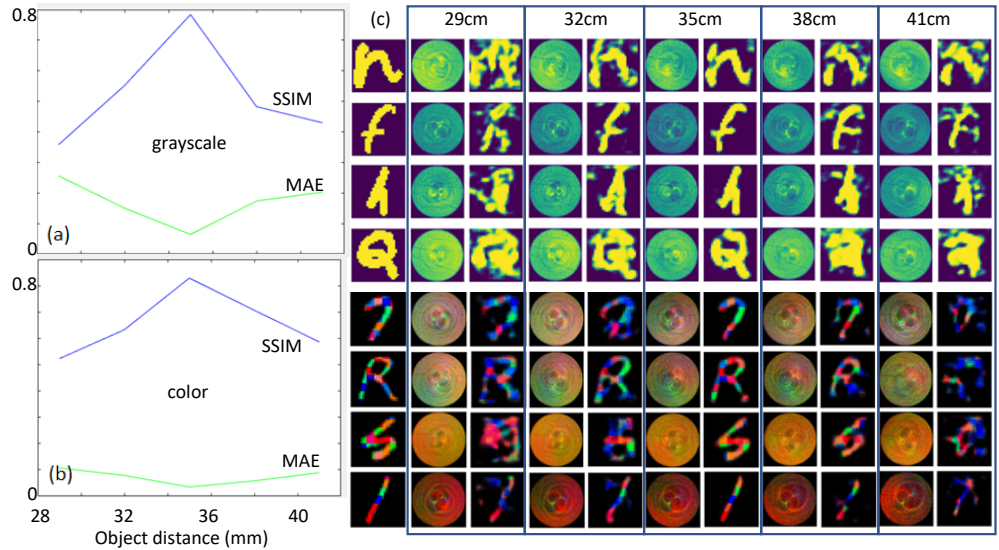

Fig. 3. Depth of focus. Average MAE and SSIM of (a) grayscale and (b) color images as function of object distance. (c) Exemplary images. Left column: ground truth. DNN trained at 35cm.

We have previously shown that an ancillary DNN can be used for predicting the image depth [17]. Following the same approach, here we trained a depth-classification network with 97,500 images captured at 5 object distances (29cm, 32cm, 35mc, 38cm and 41cm), and used 2500 images exclusively for testing. A total of 20,000 images were recorded for each object distance. The depth-classification network consists of 2D convolution blocks (one 2D convolution layer with RELU activation function followed by a batch-normalization layer) with Maxpool function between every two blocks and a final classifier[17]. The same reconstruction network as before was also trained. Exemplary results for the grayscale and color images are summarized in Fig. 4. The grayscale and color images achieved (SSIM, MAE) of (0.8, 0.06) and (0.77, 0.05), respectively. The depth prediction accuracies for the grayscale and color images were 0.9996 and 0.9988, respectively. These results suggest that with sufficient training data, it should be feasible to generate depth maps and 3D images of objects with a single frame.

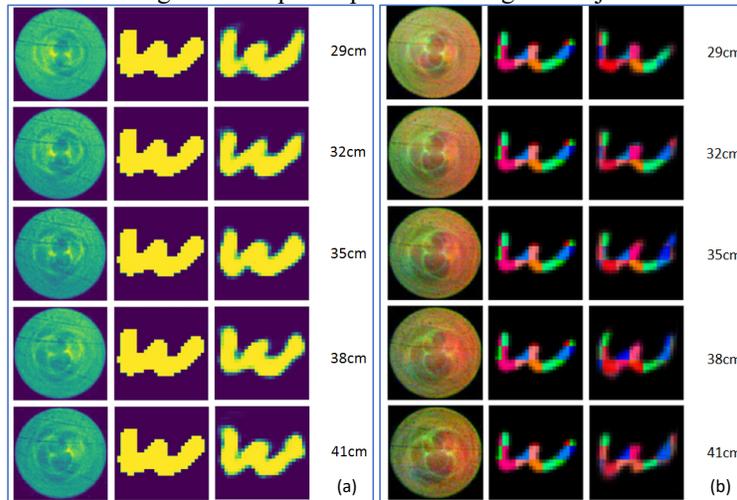

Fig. 4: Depth prediction. A separate DNN can be trained to predict the object distance. Exemplary (a) grayscale and (b) color images with predicted depths (right columns) are shown (accuracy > 99%).

Next, in order to explore the resolution of our camera, we created a new dataset (20,000 images, 1000 of which were used exclusively for testing) comprised of randomly positioned squares, whose physical size was 2.6mm. For smaller squares, the sensor signal was too weak. The DNN was able to achieve average MAE and SSIM of 0.91 and 0.01, respectively (example in Fig. 5a). At an object distance of 35cm, this corresponds to an angular resolution of ~$0.4^0$. In order to estimate the field of view (FOV) of the camera, we created a new dataset comprised of objects of total size 21cm on the LCD (randomly positioned individual squares of size 3.6mm; see Fig. 5b; 30,000 total images and 800 used exclusively for testing), and confirmed that objects within a circle of radius 5.5cm (red circle in Fig. 5b) were reconstructed with good fidelity. Since the object distance was 35cm, this corresponds to FOV of $18^0$. Similar results were obtained with color images as well.

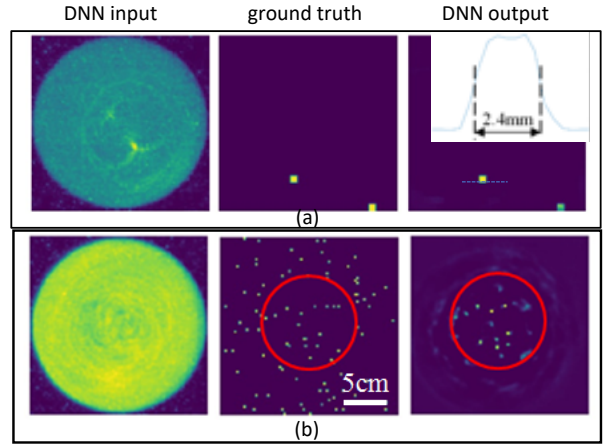

**Fig. 5.** (a) Estimating resolution by reconstructing squares as small as 2.6mm, corresponding to an angular resolution of ~$0.4^0$. Inset shows cross-section through yellow line. (b) The field of view is estimated as $18^0$ represented by the clear reconstructions within the red circle. The object distance was 35cm and similar results were obtained with color images as well.

Lastly, we show that a deep neural network can perform classification on the raw data without necessarily performing the reconstructions described above. We used two datasets, the modified EMNIST dataset from above, and 20,000 monochrome EMNIST images, each of which contains 47 classes. In both cases, 1000 images were exclusively used for testing. For classification, we used the SimpNet [22] architecture, which has shown to have better results than other state of the art networks on many benchmarks, while having much fewer parameters. For all models we use the Adam optimizer and trained for 35 epochs. The results show that the network performs better when the images are reconstructed prior to classification as opposed to classifying from the raw sensor images. In addition, the networks are more successful with the monochrome datasets as opposed to the color sets even with half the number of training images.

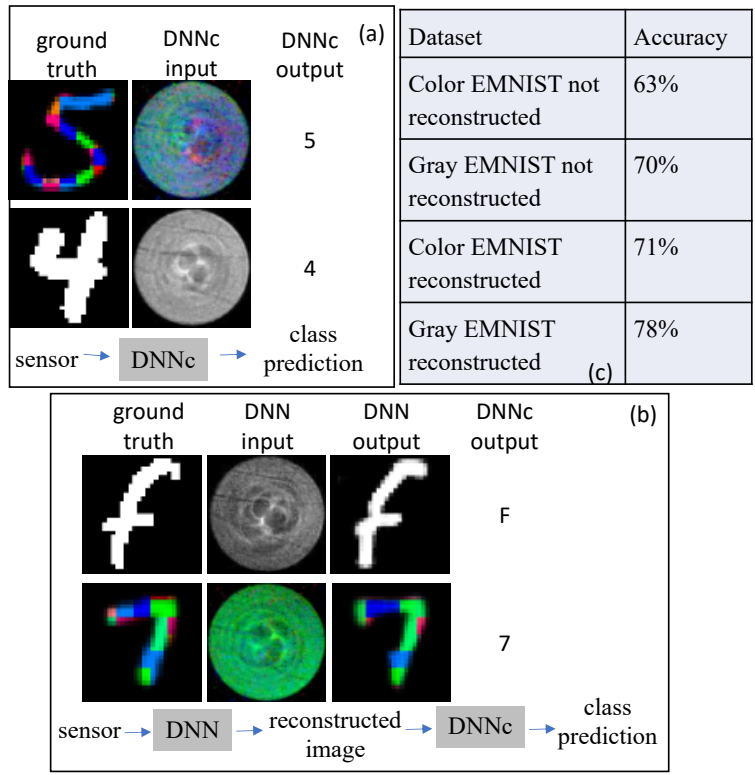

Fig. 6: Examples of classification (a) without and (b) with image reconstructions. Schematics at the bottom describe the data flow. (c) Summary of classification accuracies with 47 classes.

In this paper, we demonstrated a novel camera whose primary imaging element is a cannula of length 12.5cm and diameter=0.22mm, placed 35cm away from an object. Images were reconstructed using trained deep neural networks showcasing color, field of view of $18^0$, angular resolution of ~$0.4^0$ and depth of field of ~2cm. We further showed the potential for 3D or depth imaging. Finally, we also showed that image classification is possible not only from the reconstructed images, but also from the raw images without reconstruction. The latter is particularly interesting for application-specific imaging, where power efficiency is important and also for enhanced privacy.

## Funding, acknowledgments, and disclosures

*Funding*
National Science Foundation (1533611).
National Institutes of Health (1R21EY030717).
University of Utah Undergraduate Research Opportunities Program (UROP).

*Acknowledgments*
We thank Zhimeng Pan for discussions.

*Disclosures*
RM: University of Utah (P).

## References

1. A. Torralba and W. T. Freeman, "Accidental pinhole and pinspeck cameras," IEEE Conf. on Computer Vision and Pattern Recognition (CVPR) June 2012.
2. A. Ozcan and E. McLeod, "Lensless imaging and sensing," *Ann. Rev. Biomed. Eng.* 18(1) 77-102 (2016).
3. N. George, "Lensless electronic imaging," *Opt. Commun.* 133, 22-26 (1997).
4. G. Kim, K. Isaacson, R. Palmer and R. Menon, "Lensless photography with only an image sensor," *Appl. Opt. 56(23),6450-6456 (2017).*
5. G. Kim and R. Menon, "Computational imaging enables a "see-through" lensless camera," *Opt. Exp.* 26(18) 22826-22836 (2018).
6. S. Nelson, E. Scullion & R. Menon, "Optics-free imaging of complex, non-sparse QR-codes with Deep Neural Networks," *OSA Continuum* 3(9) 2423-2428 (2020).
7. G. Kim, S. Kapetanovic, R. Palmer and R. Menon, "Lensless-camera based machine learning for image classification," *arXiv:1709.00408 (2017).*
8. Z. Pan, B. Rodriguez & R. Menon, "Machine-learning enables Image Reconstruction and Classification in a "see-through" camera," *OSA Continuum* 3(3) 401-409 (2020)
9. G. Barbastathis, A. Ozcan and G. Situ, "On the use of deep learning for computational imaging," *Optica* 6(8) 921-943 (2019).
10. I. Papadopuolos, S. Farahi, C. Moser and D. Psaltis, "Focusing and scanning light through a multimode optical fiber using digital phase conjugation," *Opt. Exp.* 20(10) 10583-10590 (2012).
11. O. Katz, P. Heidmann, M. Fink and S. Gigan, "Non-invasive single-shot imaging through scattering layers and around corners via speckle correlations," *Nat. Photon.* 8, 784-790 (2014).
12. G. Kim and R. Menon, "An ultra-small 3D computational microscope," *Appl. Phys. Lett.* **105** 061114 (2014).
13. G. Kim, N. Nagarajan, M. Capecchi and R. Menon, "Cannula-based computational fluorescence microscopy," *Appl. Phys. Lett.* 106, 261111 (2015).
14. G. Kim and R. Menon, "Numerical analysis of computational cannula microscopy," *Appl. Opt.* 56(9), D1-D7 (2017)
15. G. Kim, N. Nagarajan, E. Pastuzyn, K. Jenks, M. Capecchi, J. Sheperd and R. Menon,"Deep-brain imaging via epi-fluorescence computational cannula microscopy," *Scientific Reports*, 7:44791 DOI: 10.1038/srep44791 (2016).
16. R. Guo, Z. Pan, A. Taibi, J. Sheperd and R. Menon, "Computational Cannula Microscopy of neurons using neural networks," *Opt. Lett.* 45(7) 2111-2114 (2020)
17. R. Guo, Z. Pan, A. Taibi, J. Shepherd and R. Menon, "3D Computational Cannula Fluorescence Microscopy enabled by artificial neural networks," *Opt. Exp.* 28(22) 32342-32348 (2020).
18. H. Lin and S. Jegelka, "ResNet with one-neuron hidden layers is a Universal approximator," Adv. in Neural Information Processing Systems (NIPS), 31, 6169-6178 (2018).
19. G. Cohen, S. Afshar, J. Tapson, and A. van Schaik, "EMNIST: an extension of MNIST to handwritten letters," http://arxiv.org/abs/1702.05373 (2017).
20. Emoji dataset from https://emojipedia.org/
21. T. Clanuwat, M. Bober-Irizar, A. Kitamoto, A. Lamb, K. Yamamoto, and D. Ha, "Deep learning for classical japanese literature," arXiv preprint arXiv: 1812.01718v1 (2018).
22. S. H. Hasanpour, M. Rouhani, M. Fayyaz, M. Sabokrou, and E. Adeli, "Towards Principled Design of Deep Convolutional Networks: Introducing SimpNet," arXiv:1802.06205 (2018).